\documentclass[aps,prb,showpacs,manuscript,floatfix]{revtex4}
\usepackage{graphicx}

\newcommand{\GVB}{$G(V,B)$}
\newcommand{\micro}[1]{$\mu\text{#1}$}
\newcommand{\fig}[1]{Fig.~\ref{#1}}

\newcommand{\eq}[1]{Eq.~(\ref{#1})}

\begin{document}

\date{\today}

\author{Yaroslav Tserkovnyak}
\author{Bertrand I. Halperin}
\affiliation{Lyman Laboratory of Physics, Harvard University,
Cambridge, Massachusetts 02138, USA}
\author{Ophir M. Auslaender}
\author{Amir Yacoby}
\affiliation{Dept. of Condensed Matter Physics, Weizmann
Institute of Science, Rehovot 76100, Israel}

\title{Signatures of spin-charge separation in double--quantum-wire tunneling}

\begin{abstract}
We present evidence for spin-charge separation in the tunneling spectrum
of a system consisting of two quantum
wires connected by a long narrow tunnel junction at the edge of
a GaAs/AlGaAs bilayer heterostructure. Multiple excitation velocities are
detected in the system by tracing out electron spectral peaks in the
conductance dependence on the applied voltage, governing the energy
of tunneled electrons, and the magnetic field, governing the momentum
shift along the wires. The boundaries of the
wires are important and lead to a characteristic interference
pattern in measurements on short junctions. We show that the
experimentally observed modulation of the conductance
oscillation amplitude as a function of the voltage bias can also
be accounted for by spin-charge separation of the
elementary excitations in the interacting wires.

Manuscript for proceedings of the NATO workshop on Theory of Quantum Transport
in Metallic and Hybrid Nanostructures, St. Petersburg, August 2003, to be
published by Kluwer Academic Publishers B.V., Dordrecht, The Netherlands,
edited by V. Kozub and V. Vinokur. Talk was presented by B. I. Halperin.
\end{abstract}


\maketitle

\section{Introduction}

One-dimensional (1D) electronic systems are a very fertile ground for studying the physics of interacting many-body systems. Gapless electron gases in one dimension possess universal low-energy properties which can be mapped onto the exact solution of the Luttinger model.\cite{Voit:rpp94} Such 1D systems are collectively termed as Luttinger liquids (LL's). Despite a number of remarkable predictions for electronic and thermodynamic properties of LL's,\cite{Voit:rpp94} some made more than twenty years ago, direct experimental verification for many of them has remained a challenge. This mostly owes to the high quality of quasi-1D systems necessary to bridge the gap between the usual Landau Fermi-liquid (FL) physics in three spatial dimensions and the LL physics in one dimension.

An important prediction of LL theory is that the low-energy elementary excitations of a one-dimensional metal are not electronic quasiparticles, as in the Landau FL theory of three-dimensional Fermi systems, but rather are separate spin and charge excitations that propagate at different velocities. An electron entering an LL will split into spin and charge excitations, and the electron propagator will have singularities corresponding to both velocities, in contrast to the case of a Landau FL where there is a simple pole at a single Fermi velocity. In this paper, we discuss evidence for spin-charge separation in tunneling between two parallel quantum wires at a cleaved edge of a double--quantum-well heterostructure. We use two approaches: one based on mapping out the elementary-excitation dispersions by measuring the conductance $G$ as a function of the magnetic field $B$ applied perpendicular to the plane connecting the wires and the voltage bias $V$, and the other focusing on the conductance oscillation pattern, in the $(V,B)$ plane, arising due to the finite length of the tunnel junction.

\section{Experimental Method}

\begin{figure}
\includegraphics[width=0.8\linewidth,clip=]{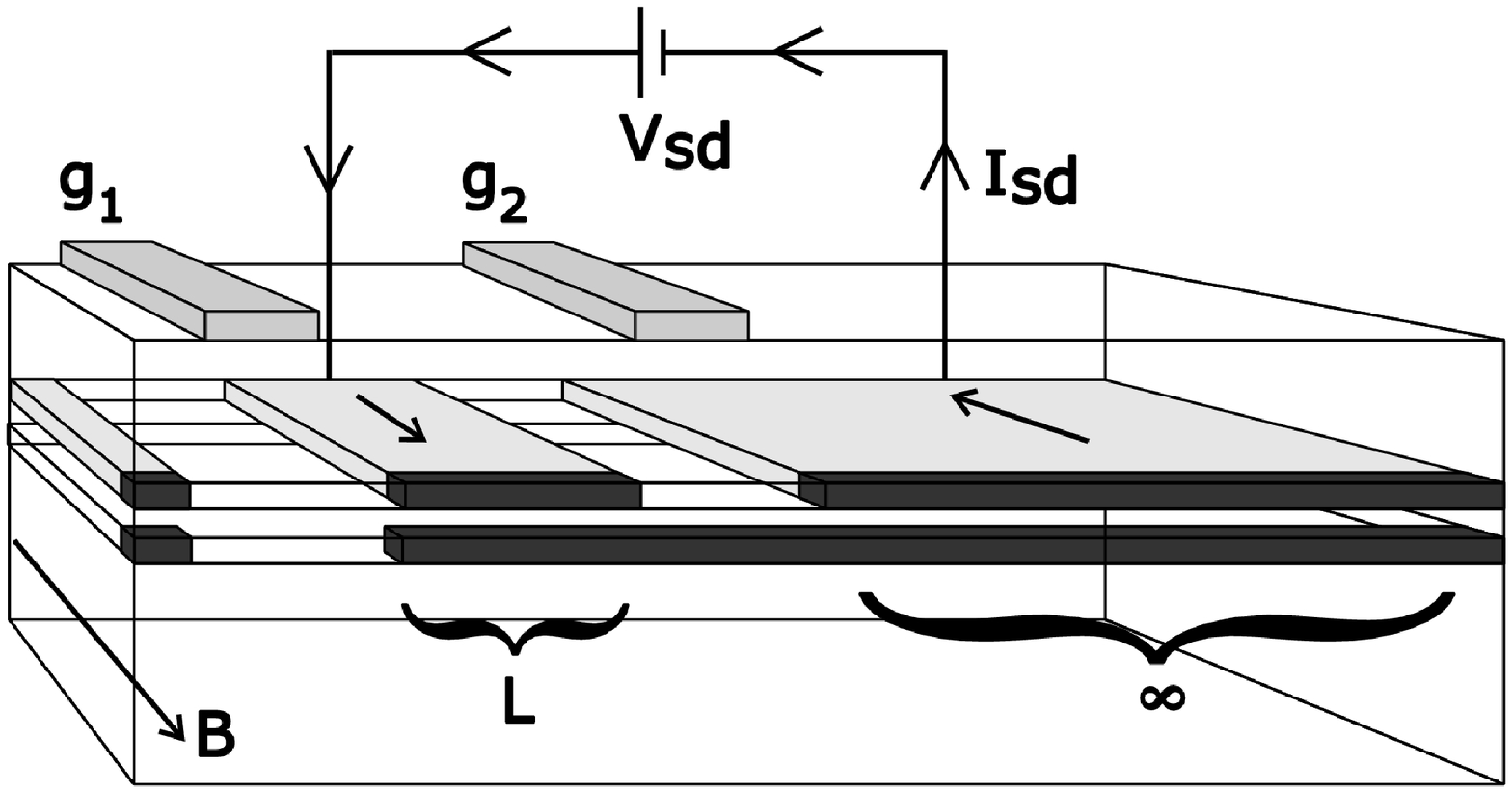}
\caption{\label{circuit}Illustration of the sample and the
contacting scheme. The sample is fabricated using the CEO
method. The parallel 1D wires
span along the whole cleaved edge (front side in the
schematic). The upper wire (UW) overlaps the 2DEG, while the lower
wire (LW) is separated from them by a thin AlGaAs barrier
(the wires are shown in dark gray and the 2DEG is light gray).
Contacts to the wires are made through the 2DEG. Several tungsten top
gates can be biased to deplete the electrons under them: We
show only $g_1$, here biased to deplete the 2DEG and both
wires, and $g_2$, here biased to deplete only the 2DEG and
the upper wire. The magnetic field $B$ is perpendicular to
the plane defined by the wires. The depicted configuration
allows the study of the conductance of a tunnel junction
between a section of length $L$ of the upper wire and a
semi-infinite lower wire.}
\end{figure}

The two parallel 1D wires are fabricated by cleaved-edge
overgrowth (CEO), see \fig{circuit}.
Initially, a GaAs/AlGaAs heterostructure with two closely situated
parallel quantum wells is grown. The upper quantum well is 20~nm
wide, the lower one is 30~nm wide and they are separated by a 6~nm
AlGaAs barrier about 300~meV high. We use a modulation doping
sequence that renders only the upper quantum well occupied by a
two-dimensional electron gas (2DEG) with a density
$n\approx2\times10^{11}~\text{cm}^{-2}$ and mobility
$\mu\approx3\times10^6~\text{cm}^2\text{V}^{-1}\text{s}^{-1}$. After cleaving
the sample in the molecular beam epitaxy growth chamber and
growing a second modulation doping sequence on the cleaved edge,
two parallel quantum
wires are formed in the quantum wells along the whole side of the
sample. Both wires are tightly confined on three sides by
atomically smooth planes and on the fourth side by the triangular
potential formed at the cleaved edge.

Spanning across the sample are several tungsten top gates of width
2~\micro{m} that lie 2~\micro{m}\ from each other (two of these
are depicted in \fig{circuit}). The differential conductance $G$
of the wires is measured through indium contacts to the 2DEG
straddling the top gates. While monitoring $G$ with standard lock-in
techniques (we use an excitation of $10$~\micro{V} at $14$~Hz) at
$T=0.25$~K, we decrease the density of the electrons under a
gate by decreasing the voltage on it ($V_g$).
At $V_g=V_{\text{2D}}$, the 2DEG depletes and $G$
drops sharply, because the electrons have to scatter into the
wires in order to pass under the gate. For $V_{\text{2D}}>V_g>V_U$
the conductance drops stepwise each time a mode in the upper
wire is depleted.\cite{Yacoby:prl96} In this voltage range, the
contribution of the lower wire to $G$ is negligible because it is
separated from the upper quantum well by a tunnel barrier.
When $V_g=V_U$, the upper wire depletes and only the lower wire
can carry electrons under the gate. This last conduction channel
finally depletes at $V_L$ and $G$ is suppressed to zero.

The measurements are performed in the configuration depicted
in \fig{circuit}. The source is the 2DEG between two gates,
$g_1$ and $g_2$ in \fig{circuit}, the voltages on which
are $V_1<V_L$ and $V_L<V_2<V_U$, respectively. The upper wire
between these gates is at electrochemical equilibrium with
the source 2DEG. This side of the circuit is separated by the
tunnel junction we wish to study from the drain. The drain is
the 2DEG to the right of $g_2$ (the semi-infinite 2DEG in
\fig{circuit}) and it is in equilibrium with the right,
semi-infinite, upper wire and with the whole semi-infinite
lower wire in \fig{circuit}. Thus, any voltage difference ($V$)
induced between the source and the drain drops on the
narrow tunnel junction between the gates. In addition,
we can shift the momentum of the tunneling electrons with a magnetic
field $B$ perpendicular to the plane defined by the wires. This configuration
therefore gives us control over both the energy and the momentum of the
tunneling electrons.

\section{Dispersions of Elementary Excitations}
\label{dis}

\begin{figure}
\includegraphics[width=0.9\linewidth,clip=]{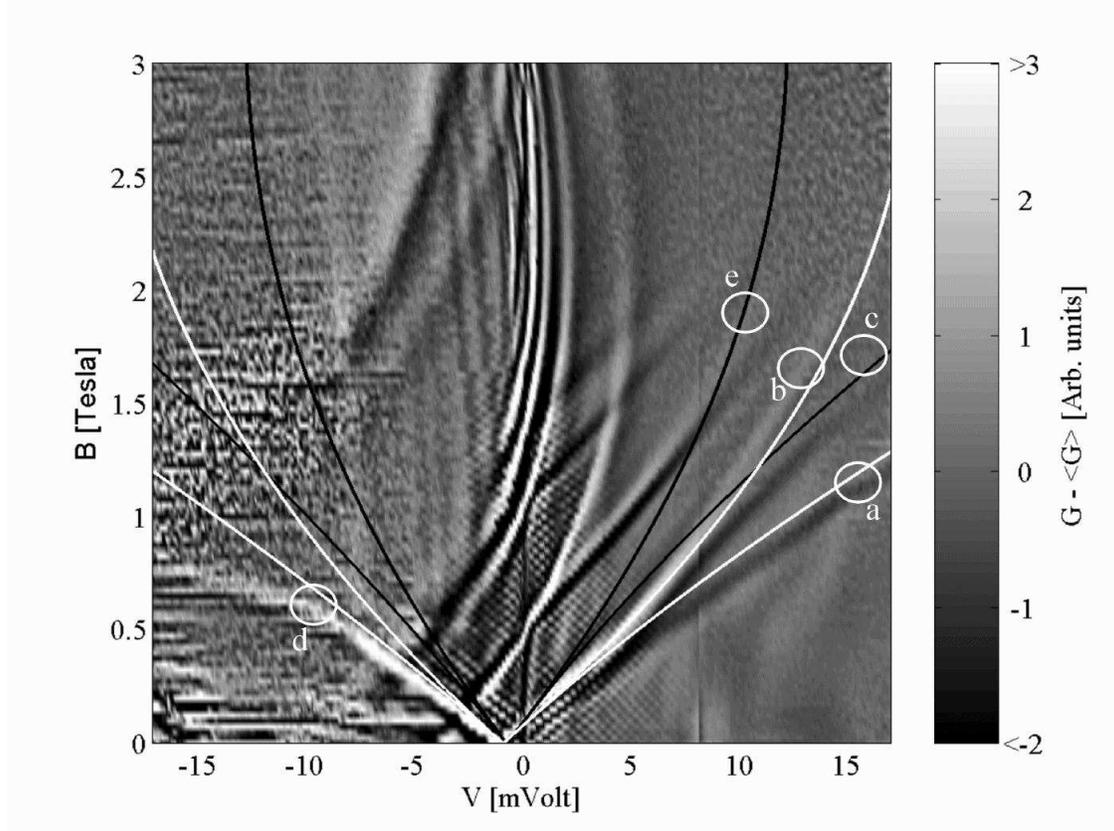}
\caption{\label{GVB}Measurement of \GVB\ for a 2~\micro{m}
junction. Light shows positive and dark negative differential conductance.
A smoothed background has been subtracted to emphasize the spectral peaks
and the finite-size oscillations. The solid black lines are the expected dispersions of noninteracting electrons at the same electron densities as the lowest 1D bands of the wires, $\left|u_1\right>$ and $\left|l_1\right>$. The white lines are generated in a similar way but after rescaling the GaAs band-structure mass, and correspondingly the low-voltage slopes, by a factor of $0.7$. Only the lines labeled by $a$, $b$, $c$, and $d$ in the plot are found to trace out the visible peaks in \GVB, with the line $d$ following the measured peak only at $V>-10$~mV.}
\end{figure}

The conductance for a spacing of 2~\micro{m} between gates $g_1$ and $g_2$ is shown in \fig{GVB}. The measured bright and dark curves in the plot can be interpreted as spectral peaks tracing out the dispersions of the elementary excitations in the wires.\cite{Auslaender:sc02} In the case of noninteracting electrons, the curves are expected to map out parabolas defining the continua of electron-hole excitations across the tunnel barrier for various pairs of 1D modes, one in the UW and the other in the LW. At small voltages, electron repulsion is predicted to split the curves into branches with slopes corresponding to different charge- and spin-excitation velocities, crossing at a point with $V=0$ and magnetic field necessary to compensate for the Fermi wave-vector mismatch between the 1D modes (in the following referred to as the ``crossing point'').\cite{Carpentier:prb02,Zulicke:prb02}

The 1D modes in the upper quantum well are coupled to the 2DEG
via elastic scattering which ensures an Ohmic contact
between the 2D well states and the
states confined to the cleaved edge.\cite{Picciotto:prl00}
(This scattering is however weak on the scale of the junction length,
not affecting the finite-size quantum-interference effects in
tunneling.\cite{Tserkovnyak:prb032})
In addition to tunneling between the 1D states of the wires, there is
a direct electron transfer from the 2DEG to the lower wire, when the
extended states have an appreciable weight on the edge. Each of
the quasi-1D wires carries several 1D modes. In our theoretical
analysis, we will only consider the transition between
the lowest 1D bands of the wires (i.e., the bands with the
largest Fermi momentum), $\left|u_1\right>\leftrightarrow\left|l_1\right>$, which has a strong signal, as seen in \fig{GVB}, given by a family of curves crossing at the lowest magnetic field. Since other 1D modes have Fermi velocities smaller than $\left|u_1\right>$ and $\left|l_1\right>$ by at least 40~\%,\cite{Auslaender:sc02} we can disregard their coupling to the lowest bands,\cite{Matveev:prl93} keeping in mind, however, that the effective electron-electron interaction is affected by screening due to the nearby 2DEG and other 1D modes. Both the spin-orbit interaction and Zeeman splitting are negligible in comparison to the Fermi energy, so that the electron states are nearly spin degenerate in our heterostructure. In our theoretical discussion we therefore consider tunneling between two coupled spinful modes having some effective intrawire and interwire interaction. To this end, we use Luttinger-liquid formalism,\cite{Voit:rpp94} assuming sufficiently low temperature and voltage bias. It is important in our analysis that the measured densities of $\left|u_1\right>$ in the UW and $\left|l_1\right>$ in the LW happen to match to within several per cent,\cite{Auslaender:sc02} so that the electronic excitations in the double-wire system are collective across the tunnel barrier as well as within each wire.

\begin{figure}
\includegraphics[width=0.8\linewidth,clip=]{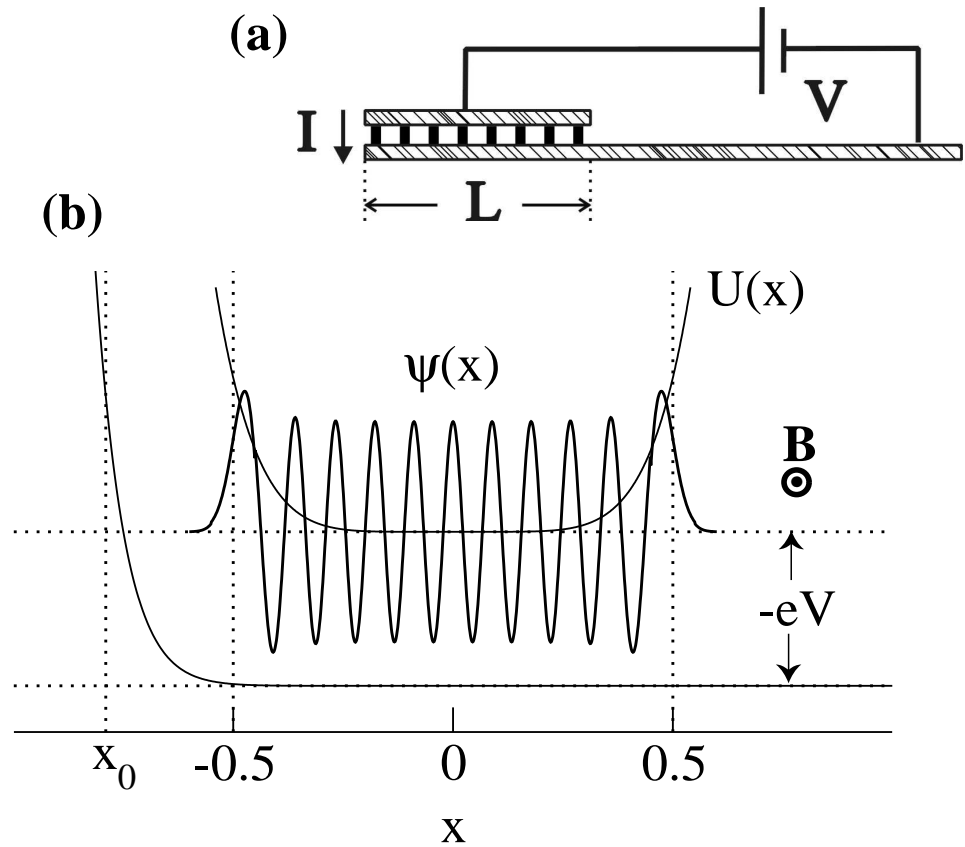}
\caption{\label{model}Schematics of the circuit (a) and the
model (b). A wire of length $L$ runs parallel to a
semi-infinite wire. The boundary of the upper wire is
formed by potential $U(x)$ confining the one-electron wave
function $\psi(x)$ along the wire.
The energy and momentum of the tunneling electrons are governed
by applied voltage $V$ and magnetic field $B$.}
\end{figure}

The geometry for our theoretical description is shown in
\fig{model}. The potential well $U(x)$ is
felt by electrons in the upper quantum
wire, which are confined to a region
of finite length by potential gates at both ends (see the
source region in \fig{circuit}). One of these gates ($g_1$)
causes the electrons in the lower wire to be reflected at one end,
but the other ($g_2$) allows them to pass freely under it. The
effective tunneling region is determined by the length of the
upper wire, which is approximately the region $|x| < L/2$ in
\fig{model}. The magnetic field, $B$, gives a momentum boost
$\hbar q_B=eBd$ along the \textit{x}-axis to the electrons
tunneling from the upper to the lower wire.
The current (for electrons with a given spin)
\begin{equation}
I=e|\lambda|^2\int_{-\infty}^\infty dxdx^\prime
\int_{-\infty}^\infty dte^{iq_B(x-x^\prime)}e^{ieVt/\hbar}C(x,x^\prime;t)\,,
\label{Ie}
\end{equation}
is determined to lowest order in perturbation theory by the two-point Green function\cite{Carpentier:prb02}
\begin{eqnarray}
C(x,x^\prime;t)&=&\left\langle\left[\Psi^\dagger_l\Psi_u(x,t),\Psi^\dagger_u\Psi_l(x^\prime,0)\right]\right\rangle\nonumber\\&=&G^>_u(x,t;x^\prime,0)G^<_l(x^\prime,0;x,t)-G^<_u(x,t;x^\prime,0)G^>_l(x^\prime,0;x,t)\,.
\label{C}
\end{eqnarray}
The last equality in \eq{C} is valid when the interwire electron-electron interactions vanish. Although it might not be a good approximation for our closely-spaced wires, for pedagogical reasons we will discuss this limit first. One-particle correlators are defined by the usual expressions: $G^>(x,t;x^\prime,t^\prime)=-i\left\langle\Psi(x,t)\Psi^\dagger(x^\prime,t^\prime)\right\rangle$ and $G^<(x,t;x^\prime,t^\prime)=i\left\langle\Psi^\dagger(x^\prime,t^\prime)\Psi(x,t)\right\rangle$. For $V>0$ ($V<0$), only the $G^>_uG^<_l$ ($G^<_uG^>_l$) term in \eq{C} contributes to the current (\ref{Ie}).

In the LL picture, we can distinguish between the left- and right-moving electronic excitations in a given 1D mode.\cite{Voit:rpp94} In long wires, each chirality contributes terms proportional to $e^{\pm ik_F(x-x^\prime)}$ to the one-particle Green functions (away from the boundaries), where $k_F=(\pi/2)n$ and $n$ is the electron density in the mode. If the magnetic field is small enough, $q_B\ll k_F$, the edge-state chirality of the electrons cannot be changed during a tunneling event, and the total current is thus a sum of the right-moving and left-moving contributions.  It is sufficient to calculate the tunneling rate of the right movers only, since it equals to that of the left movers under the magnetic-field reversal, $B\rightarrow-B$ (so that the total conductance is an even function of $B$). The corresponding zero-temperature one-particle Green functions for a gapless translation- and spin--rotation-invariant 1D gas of interacting electrons has a universal form
\begin{eqnarray}
G^{>,<}(x,t;x^\prime,0)&=&\frac{\pm\psi(x)\psi^\ast(x^\prime)}{2\pi\sqrt{(z-v_st\pm i0^{+})(z-v_ct\pm i0^{+})}}\left[\frac{r_c}{\sqrt{z^2-(v_ct\mp ir_c)^2}}\right]^{\alpha}\,,
\label{ggl}
\end{eqnarray}
where $v_c=v_F/K$ is the charge-excitation velocity, which is enhanced with respect to the Fermi velocity $v_F$ by electron repulsion, $v_s$ is the spin-excitation velocity, which is close to $v_F$ for vanishing backscattering rate and is determined by the exchange interaction of neighboring electrons for strong repulsion with a sizable backscattering, and $\alpha=(K+K^{-1}-2)/4$ is a nonuniversal exponent. $K<1$ is the compressibility normalized to that of the free-electron gas at the same density, $z=x-x^\prime$, and $r_c$ is a short-distance cutoff. $\psi(x)=e^{ik_Fx}$ is the noninteracting-electron wave function for an infinite wire at the Fermi level. Using \eq{ggl} to calculate the two-particle Green function (\ref{C}) and then performing the integration in \eq{Ie}, one shows that the interactions do not shift the position of the crossing point at $V=0$ and $q_B=\Delta k_F$, the mismatch in the Fermi wave vectors of the wires, but are manifested by multiple peaks in the tunneling conductance \GVB, which intersect at the crossing point with slopes determined by different spin and charge velocities.\cite{Carpentier:prb02,Zulicke:prb02}

It turns out that in a symmetric double-wire structure, the interwire electron-electron interactions, do not change the two-point correlation function (\ref{C}) apart from renormalizing the parameters entering \eq{ggl}.\cite{Zulicke:prb02} The reason for this is that a tunneling event from the UW to the LW at a low magnetic field creates a long-lived exciton across the tunnel-barrier with the electron in the lower and hole in the upper wire, moving in the same direction. The exciton propagates freely, as an acoustic plasmon in a single wire but with the velocity $v_c$ reduced by the electron-hole attraction (corresponding to a larger normalized compressibility $K$).\cite{Tserkovnyak:prb032} Similarly, $v_s$ entering \eq{ggl} should be thought of as the antisymmetric spin velocity of coupled antiferromagnetic Heisenberg chains, again reducing to $v_F$ in the case of vanishing backscattering. For a symmetric double-wire structure, one therefore expects two velocities to be present at low magnetic-field and bias data: the antisymmetric spin- and charge-excitation velocities. The situation is more complicated at high magnetic fields capable of flipping the electron cleaved-edge chirality: A tunneled particle moves in the direction opposite to the hole left behind, decaying into a combination of symmetric and antisymmetric excitations across the tunnel barrier, even in the case of a perfectly symmetric double wire.\cite{Carpentier:prb02} We will not discuss this regime here.

One can see a family of measured curves crossing at $V=0$ and $B\approx0.1$~T in \fig{GVB}, which constitute the signal from the $\left|u_1\right>\leftrightarrow\left|l_1\right>$ tunneling. We also draw in \fig{GVB} as black solid lines the expected parabolic dispersions for noninteracting electrons at the same electron densities as $\left|u_1\right>$ and $\left|l_1\right>$; the white solid lines are generated in a similar way but after rescaling the GaAs band-structure mass, and correspondingly the low-voltage slopes, by a factor of $0.7$. Remarkably, for positive voltages, we can fit the three visible experimental curves, $a$, $b$, and $c$, by such parabolas crossing at $V=0$ with two different slopes. Such a fitting is a naive extrapolation of LL spectrum at small voltages, where the excitation dispersions can be linearized and the two visible low-voltage slopes can be associated with elementary 1D excitations. Understanding the high-voltage regime, where the dispersions acquire a curvature, requires going beyond LL theory, and for that matter beyond the scope of this paper. For negative voltages, fewer \GVB\ peaks are visible after we subtract a large background signal due to the direct 2DEG-1D tunneling; in particular, only one curve, $d$, is visible for the  $\left|u_1\right>\leftrightarrow\left|l_1\right>$ tunneling, which cannot be fitted by our naive procedure at $V<-10$~mV. It is important to point out that the observed curves $a$, $b$, and $c$ in \fig{GVB} rule out the noninteracting-electron picture for tunneling between two 1D modes with different Fermi velocities: If the LW had the higher Fermi velocity, we would expect to see dispersions $e$ and $a$ in \fig{GVB}, and if the UW had the higher velocity, we would see only $b$ and $c$, but not three curves.

The curve $c$ slope is given by the Fermi velocity of the noninteracting electron gas, corresponding to the value of the electron density measured in Ref.~\onlinecite{Auslaender:sc02}. The slope of $a$ has the velocity enhanced by a factor of $1.4$. Identifying the faster velocity with the charge mode and the slower with the spin mode, we arrive at the following LL parameters characterizing the system: $K\approx0.7$ and $v_s\approx v_F$. This value of $K$ indicates that the electron-electron interaction energy in the cleaved-edge quantum wires is comparable to the Fermi energy, resulting in a sizable effect on the correlation and thermodynamic properties. The closeness of the spin velocity $v_s$ to the noninteracting Fermi velocity implies a small backscattering rate due to electron repulsion, as expected in our wide quantum wires.\cite{Carpentier:prb02}

\section{Finite-Size Effects}
\label{fin}

\begin{figure}
\includegraphics[width=0.8\linewidth,clip=]{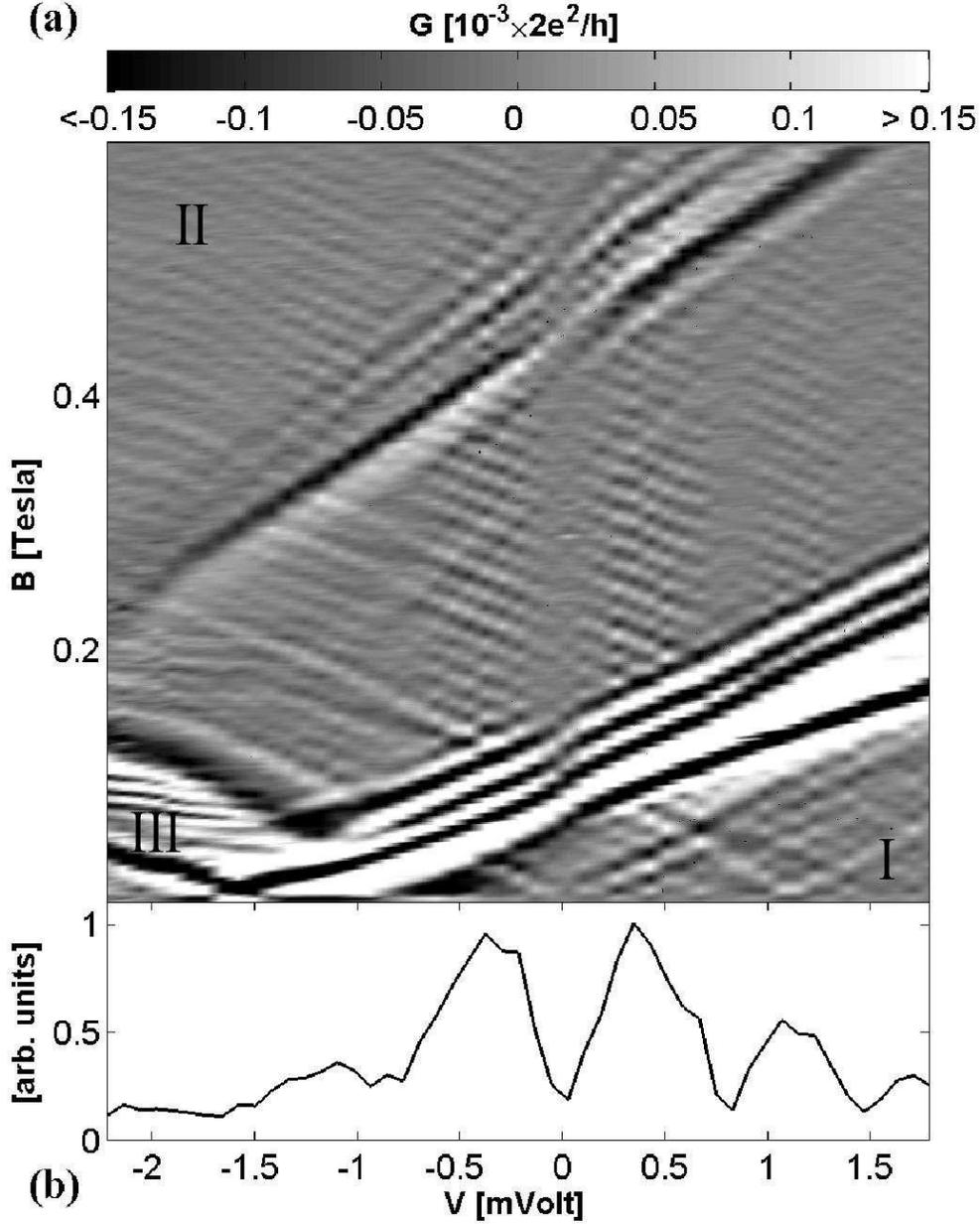}
\caption{\label{osc} Nonlinear conductance oscillations at
low field from a 6~\micro{m} junction. (a) shows the
oscillations as a function of both $B$ and $V$. (A smoothed
background has been subtracted to emphasize the
oscillations.) The brightest (and darkest) lines, corresponding to
tunneling between the lowest modes, break the $V$-$B$ plain into
regions I, II, and III. Additional positively-sloped bright and dark lines
in II arise
from other 1D channels in the wires and are disregarded in
our theoretical analysis. Also present is a slow modulation
of the strength of the oscillations along the abscissa.
(b) Absolute value of the peak of the Fourier transform
of the conductance at a fixed $V$ in region $II$ as a function of $V$.
Its slow modulation as a function of $V$ is easily discerned.}
\end{figure}

The momentum of the electrons tunneling through a window of finite length $L$ is only conserved within an uncertainty of order $2\pi/L$, resulting in conductance oscillations away from the main dispersion peaks. We show in this section that another spectacular manifestation of spin-charge separation at small voltages can be tracked down in such oscillations, as those forming checkerboard-like patterns near the crossing points of various dispersion curves in \fig{GVB}a. We zoom into these oscillations in \fig{osc}a. In order to understand them in detail, we now generalize our analysis to take into account the finite length of the upper wire.

Assuming that the electron density in each wire varies slowly on the length scale of $k_F^{-1}$ (except for unimportant regions very close to the boundaries), we use the WKB wave function
\begin{equation}
\psi(x)=\frac{e^{i k_Fx}e^{-is(x)}}{\sqrt{k(x)}}\,,
\label{WKB}
\end{equation}
where $k(x)=k_F[1-U(x)/E_F]^{1/2}$ and $s(x)=\int_0^xdx^\prime[k_F-k(x^\prime)]$, for right-moving electrons in the UW in \eq{ggl}. $U(x)$ is the potential formed by the top gates defining the finite length of the UW, see \fig{model}. The right movers in the LW are taken to be propagatory, as in the infinite wire, assuming its left boundary is formed by the gate $g_1$ in the region where the UW is already depleted and assuming the gate $g_2$ potential in the LW is well screened. Additional assumptions required for using the WKB wave function (\ref{WKB}) in the correlator of a finite interacting wire are discussed in Ref.~\onlinecite{Tserkovnyak:prb032}.

Substituting Green functions (\ref{ggl}) into integral (\ref{Ie}), we obtain for the tunneling current at $V>0$
\begin{equation}
I\propto\int_{-\infty}^\infty dxdx^\prime e^{i(q_B-k_F)(x-x^\prime)}\psi_u(x)\psi_u^\ast(x^\prime)h(x-x^\prime)\,,
\label{Ih}
\end{equation}
where
\begin{equation}
h(z)=-\int_{-\infty}^\infty dt\frac{e^{ieVt/\hbar}}{(z-v_st+i0^+)(z-v_ct+i0^+)}\left[\frac{r_c^2}{z^2-(v_ct-ir_c)^2}\right]^\alpha\,.
\label{hz}
\end{equation}
Here we have taken the two wires to have the same electron density and strength of the interactions. For not very strong interactions, $\alpha\ll1$, the last term in \eq{hz} can be disregarded, away from the regime of the zero-bias anomaly.\cite{Tserkovnyak:prb032} The integrand in \eq{hz} then has two simple poles yielding
\begin{equation}
h(z)\approx-2\pi i\frac{e^{ieVz/(\hbar v_s)}-e^{ieVz/(\hbar v_c)}}{(v_c-v_s)(z+i0^+)}\,.
\label{ha}
\end{equation}
Combining this with \eq{Ih}, we finally find the differential conductance $G=\partial I/\partial V$:
\begin{equation}
G(V,B)\propto\frac{1}{v_c-v_s}\left[\frac{1}
{v_s}|M(\kappa_s)|^2-\frac{1}{v_c}|M(\kappa_c)|^2\right]\,,
\label{gvb}
\end{equation}
where $\kappa_{s,c}=q_B+\Delta k_F+eV/(\hbar v_{s,c})$ (now including a small mismatch $\Delta k_F=k_{Fu}-k_{Fl}$ in the Fermi wave vectors of the UW and LW, respectively) and
\begin{equation}
M(\kappa)=\int dx\frac{e^{i\kappa x}e^{-is(x)}}{\sqrt{k(x)}}
\label{m}
\end{equation}
$s(x)$ and $k(x)$ being the same as in \eq{WKB}. $M(\kappa)$ can be found analytically using the stationary-phase approximation (SPA): $M(\kappa)$ is evaluated near positions $x^\pm$ ($x^+>x^-$) where $k(x^\pm)=k_F-\kappa$ and the integrand in \eq{m} has a stationary phase. In the case of a symmetric potential, $U(x)=U(-x)$ (so that, in particular, $x^-=x^+$), the SPA gives
\begin{equation}
M(\kappa)\propto\frac{\Theta(\kappa)}{\sqrt{\partial U(x^+)\partial x}}
\cos\left[\kappa x^+-s(x^+)-\pi/4\right]\,,
\label{mpm}
\end{equation}
where $\Theta(\kappa)$ is the Heaviside step function. The SPA approximation (\ref{mpm}) can be shown to diverge for small values of $\kappa$, where we have to resort to a numerical calculation of the integral in \eq{m}.\cite{Tserkovnyak:prl022} The form of \eq{mpm} shows that (1) the conductance is asymmetric in $\kappa$, vanishing for $\kappa<0$ (in the SPA approximation), (2) it oscillates in magnetic field (and similarly in voltage) with period $\Delta q_B=2\pi/x^+$, assuming $x^+$ to be a slow-varying function of $\kappa$ and that (3) it is a superposition of two oscillating patterns in the $(V,B)$ plane, the first (second) being constant-valued along line $\kappa_s={\rm const}$ ($\kappa_c={\rm const}$) and oscillating perpendicular to it, resulting in a moir\'{e} structure of $G(V,B)$. [Note that the conductance in \eq{gvb} so far only includes the right-movers' contribution. In order to get the total conductance, one has to add the piece which is mirror symmetric to \eq{gvb} around $B=0$.] If the two velocities $v_s$ and $v_c$ are not very different, there are two voltage scales characterizing the conductance oscillation pattern:
\begin{equation}
\Delta V=\frac{2\pi\hbar v_cv_s}{ex^+(v_c+v_s)}{ \rm ~~~and~~~ }\Delta V_{\text{mod}}=\frac{\pi\hbar v_cv_s}{ex^+(v_c-v_s)}\,.
\label{Vs}
\end{equation}
$\Delta V$ is the period of the ``fast'' oscillations, which would be present even in the absence of spin-charge separation, and $\Delta V_{\text{mod}}$ is the distance between consecutive minima in the oscillation power due to the moir\'{e} amplitude modulation in the voltage direction. The ratio between these two scales
\begin{equation}
\frac{\Delta V_{\text{mod}}}{\Delta V}=\frac{1}{2}~\frac{v_c+v_s}{v_c-v_s}=\frac{1}{2}~\frac{1+v_s/v_c}{1-v_s/v_c}
\label{sf}
\end{equation}
can be used to experimentally extract the ratio between the two velocities. We find
\begin{equation}
v_s/v_c=0.67\pm0.07\,,
\label{K}
\end{equation}
which is independent of the UW length $L$, while both $\Delta V$ and $\Delta V_{\text{mod}}$ scale roughly as $1/L$. This value is in agreement with the one found in Sec.~\ref{dis}.

\begin{figure}
\includegraphics[width=0.8\linewidth,clip=]{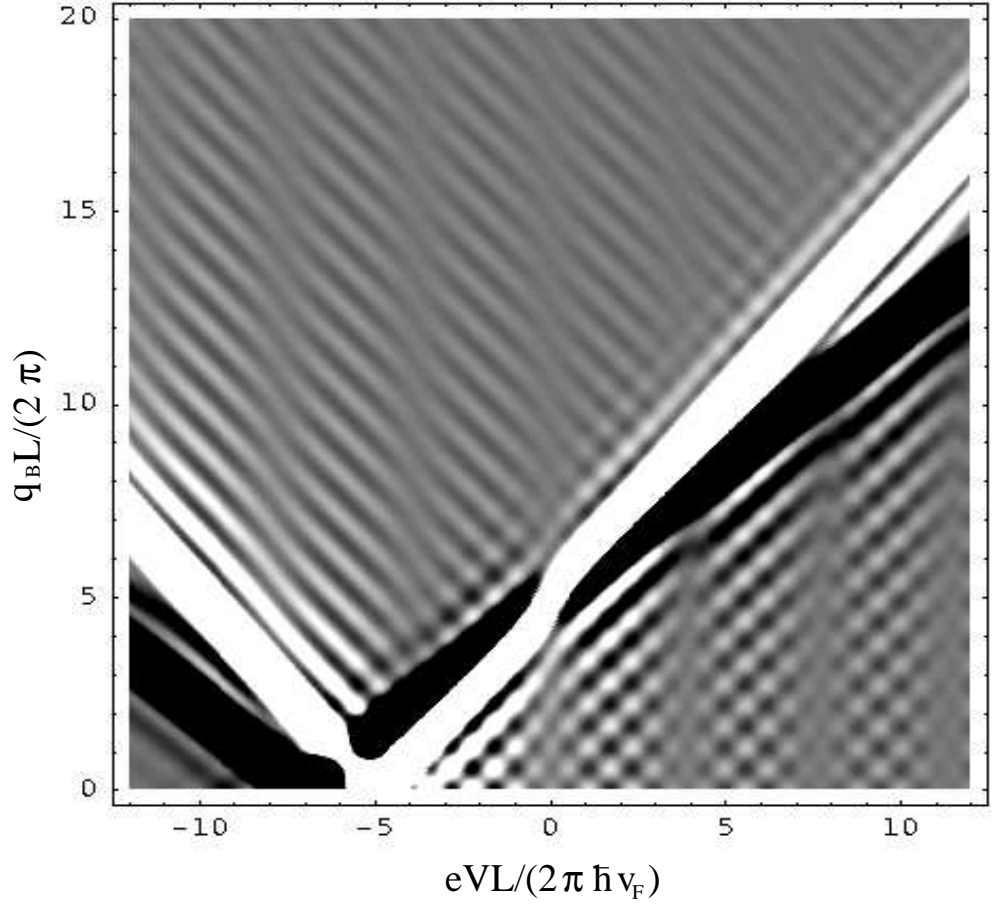}
\caption{\label{oscT}The differential conductance interference pattern
near the lower crossing point calculated using a smooth confining potential
for the upper wire. $v_c=1.4v_F$, $v_s=v_F$, and $\Delta k_F=10\pi/L$.}
\end{figure}

Finally, we compare the interference pattern predicted by our
theory, \eq{gvb}, with the experiment, Fig.~\ref{osc}(a).
$G(V,B)$ calculated using a smooth confining
potential given by $U(x)=E_F\exp[(L/2-|x|)/10]$ at the boundaries of
the upper wire is shown in \fig{oscT}. Many pronounced features
observed experimentally--the asymmetry of the side lobes, a
slow fall-off of the oscillation amplitude and period away
from the principal peaks, an interference modulation along the
$V$-axis, $\pi$ phase shifts at the oscillation suppression
stripes running parallel to the field axis--are reproduced by
the theory. There is however one experimental finding which is not captured by the presented theory: In addition to the periodic modulation of the oscillations, there is an appreciable fall-off in amplitude in the voltage direction, as can be seen in \fig{osc}(b). This dephasing can be due to the dispersion curvature which becomes appreciable with increasing voltage bias. Its discussion, however, requires going beyond the linearized LL theory, which we do not attempt here.

\section{Summary}

 The two approaches to study the size of spin-charge separation, one by mapping out the dispersions, which are independent of the tunnel-junction length, and the other based on the finite-size conductance oscillations, the frequency of which scales linearly with the junction length, are found to be in excellent agreement, giving the LL parameter $K=v_F/v_c\approx0.7$ for the antisymmetric (i.e., excitonic) collective charge excitations in the lowest modes of the double-wire structure, and $v_s\approx v_F$ for the antisymmetric spin velocity. Additional, complimentary information about the electron-electron interactions can be extracted by measuring the tunneling density-of-states exponent $\alpha$ in the regime of very small voltage bias and temperature, where the tunneling rate is suppressed as a power law (the so-called LL zero-bias anomaly).\cite{Tserkovnyak:prb032} We do not discuss this in the present paper.

\acknowledgments

This work was supported in part by the US-Israel BSF,
NSF Grant DMR 02-33773, and by a research grant from the
Fusfeld Research Fund.
YT is supported by the Harvard Society of Fellows and OMA by a grant from the
Israeli Ministry of Science.


\newpage

\end{document}